\documentclass{moriond}

\usepackage{wrapfig}

\def\Journal#1#2#3#4{{#1} {\bf #2}, #3 (#4)}

\def\PRL{\em Phys. Rev. Lett.}
\def\PRD{{\em Phys. Rev.} D}

\def\be{\begin{equation}}
\def\ee{\end{equation}}
\def\bea{\begin{eqnarray}}
\def\eea{\end{eqnarray}}

\newcommand{\Photo}{}

\begin{document}
\vspace*{4cm}
\title{Heavy-Ion and Fixed-Target Physics at LHCb}

\author{Krista Smith, for the LHCb collaboration }

\address{Nuclear and Particle Physics, Los Alamos National Laboratory, Los Alamos, NM 87545, USA}

\maketitle\abstracts{
The LHCb experiment, designed for searches of new physics in beauty and charm hadron decays, has been recording data at the Large Hadron Collider since 2010.   The physics program incorporates $p$$p$, $p$A, and AA collisions, ultra-peripheral interactions, and a unique fixed-target program. LHCb is a general purpose experiment covering the forward region, measuring particles from $p_T > 0$ at forward pseudorapidity $2<\eta<5$.  
Currently the LHCb collaboration consists of eight working groups, including the Ions and Fixed-Target (IFT) working group.  Since the Moriond QCD conference in March 2023, the IFT working group has submitted eleven physics analyses for journal publication, and here we report on five of these.}

\section{Introduction}
Since the Moriond QCD conference in March 2023, the LHCb IFT working group has submitted eleven physics analyses for journal publication.  Here we report on a selection of these in both small and large collision systems, and also provide an update on the SMOG2 fixed-target program.  Five recent LHCb analyses focus on the following collision systems: $p$$p$ collisions at $\sqrt{s_{_{NN}}}$ = 7 and 13 TeV, $p$Pb collisions at $\sqrt{s_{_{NN}}}$ = 8.16 TeV, and PbPb collisions at $\sqrt{s_{_{NN}}}$ = 5 TeV.  The results include many quarkonia-related measurements, as well as the very first charged hadron measurements of $v_2$ and $v_3$ from the LHC at forward rapidity.  We also report on baryon enhancement in high multiplicity $p$$p$ collisions, an IFT analysis selected for a Physics Feature by \textit{Physical Review Letters}.

\section{Results from $p$$p$ collision data}

\begin{figure}
\begin{minipage}{1\linewidth}
\centerline{\includegraphics[width=1\linewidth]{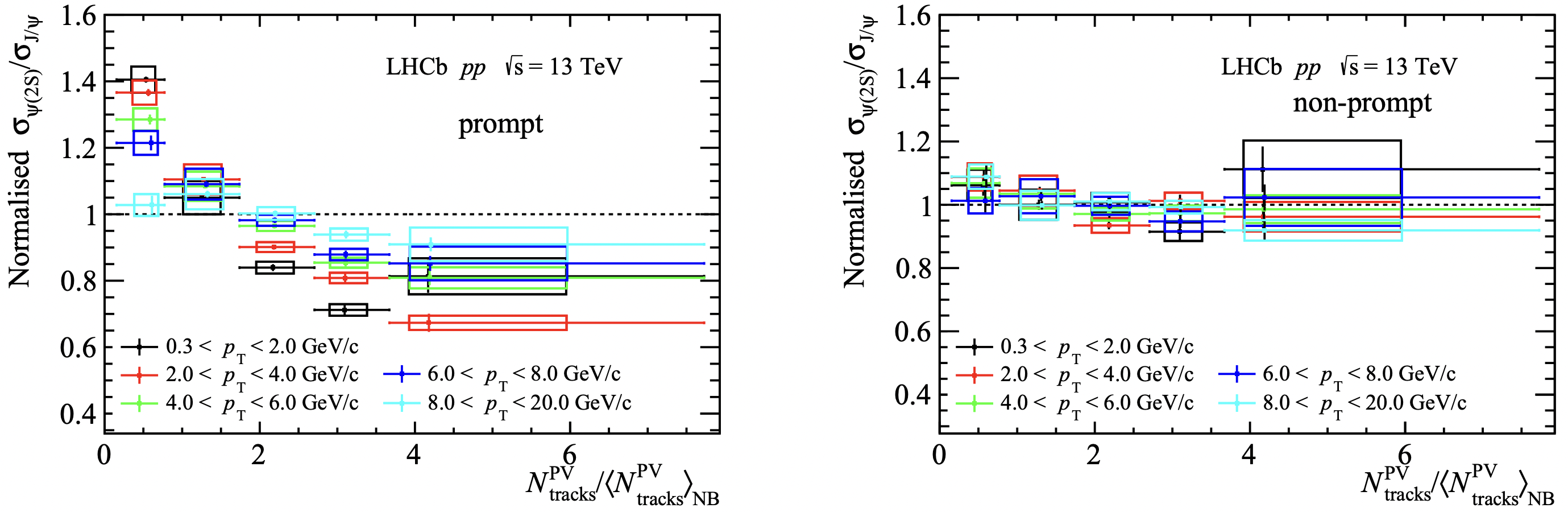}}
\end{minipage}
\caption[]{Normalised prompt (left) and nonprompt (right) $\psi(2S)$ to $J/\psi$ ratio versus multiplicity for different $p_{T}$ ranges in 13 TeV $p$$p$ collisions~\cite{results1}.  Multiplicity is measured using charged particle tracks from the primary vertex.}
\label{fig:results1}
\hfill

\begin{minipage}{0.45\linewidth}
\centerline{\includegraphics[width=1.15\linewidth]{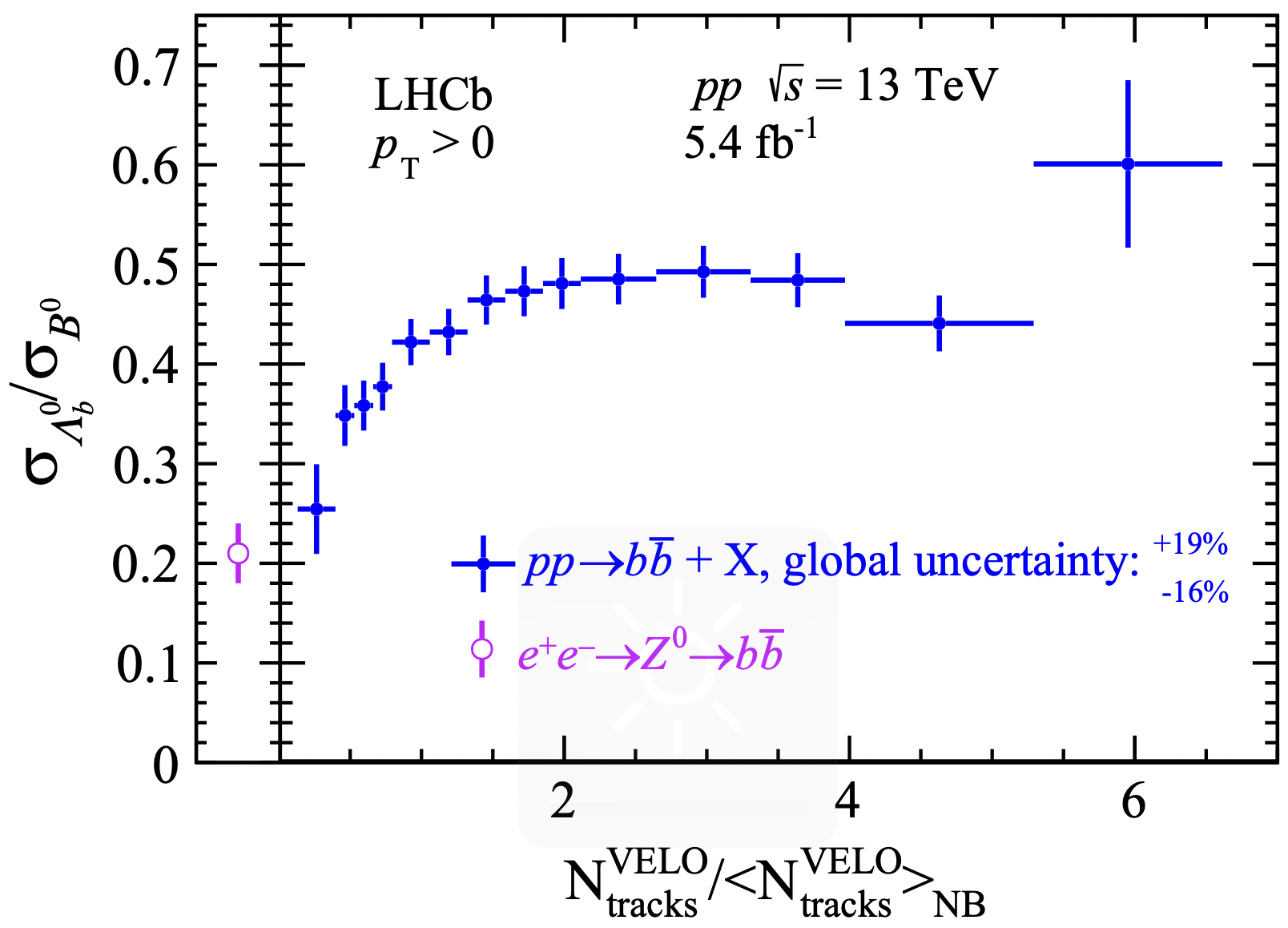}}
\end{minipage}
\hfill\begin{minipage}{0.45\linewidth}
\centerline{\includegraphics[width=0.95\linewidth]{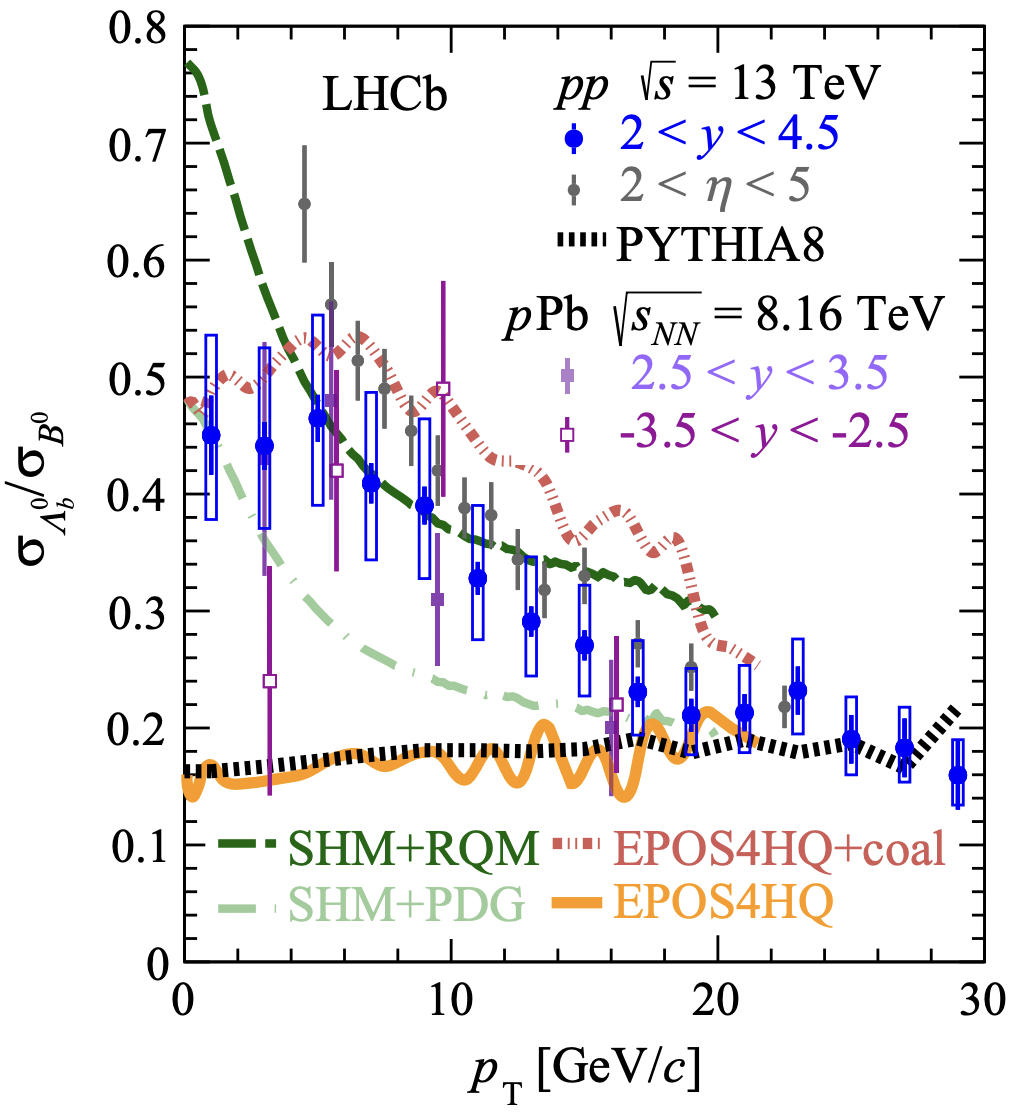}}
\end{minipage}
\caption[]{$\sigma_{\Lambda^{0}_{b}}$ to $\sigma_{B^{0}}$ ratio in $p$$p$ collisions (blue data points) at $\sqrt{s_{_{NN}}}$=13 TeV is shown as a function of normalized charged particle tracks measured by the VELO detector (left) and as a function of $p_{T}$ (right)~\cite{results2}.  }
\label{fig:results2}
\hfill

\begin{minipage}{1\linewidth}
\centerline{\includegraphics[width=0.9\linewidth]{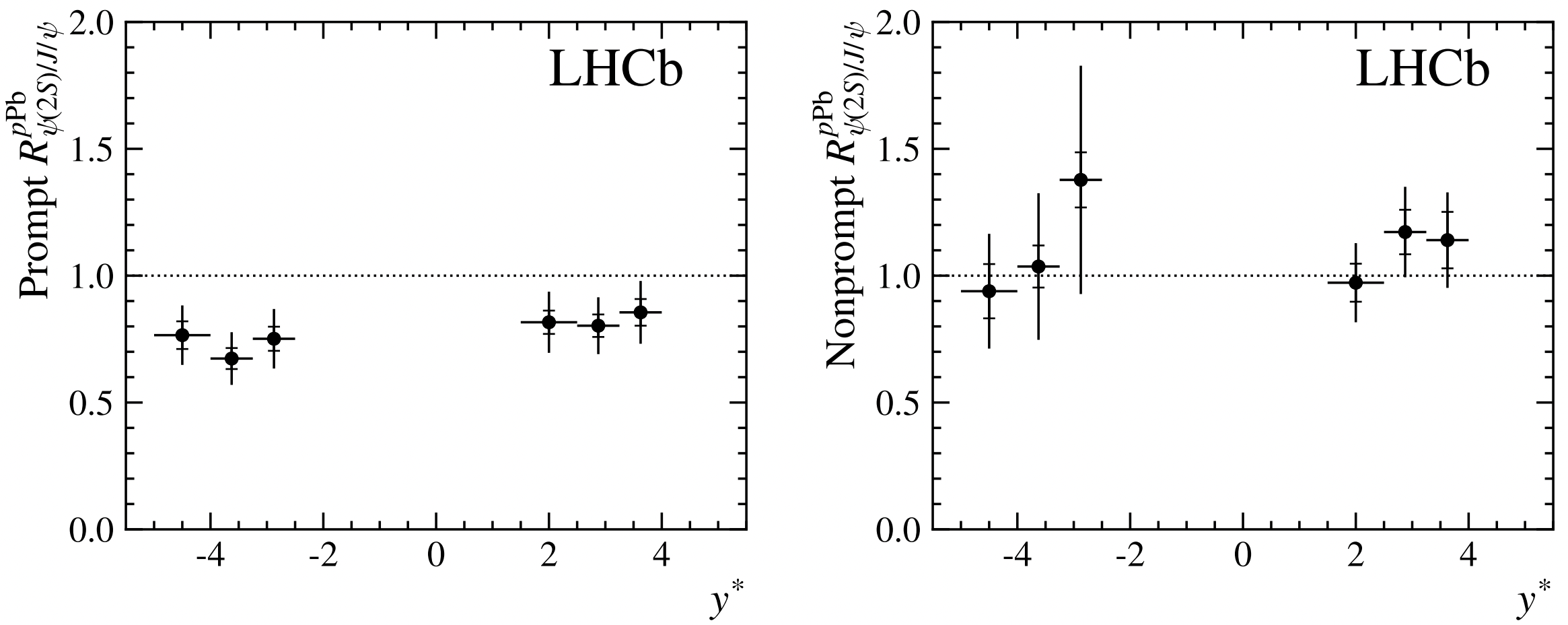}}
\end{minipage}
\caption[]{Double ratio of prompt $\psi(2S)$ to $J/\psi$ cross section as a function of rapidity for the prompt (left) and nonprompt (right) contributions~\cite{results3}.  The double ratio compares the $\psi(2S)$ to $J/\psi$ ratio in $p$Pb collisions with the same measurement in $p$$p$ collisions.}
\label{fig:results3}
\end{figure}

The normalised prompt and nonprompt $\psi(2S)$ to $J/\psi$ ratio versus multiplicity is shown in Figure~\ref{fig:results1} for various $p_T$ ranges in 13 TeV $p$$p$ collisions~\cite{results1}. Multiplicity is measured using charged particle tracks from the primary vertex (PV tracks).  When taking the ratio of the excited to ground charmonium state, many systematic uncertainties cancel.  Additionally, initial state effects are also expected to largely cancel.  The nonprompt production (right) is fairly flat, showing little multiplicity dependence.  However, the suppression observed with increasing multiplicity at low $p_T$ (red and black data points) for prompt production only (left) is consistent with final state effects in $p$$p$ collisions.   

\begin{figure}
\begin{minipage}{0.45\linewidth}
\centerline{\includegraphics[width=1.15\linewidth]{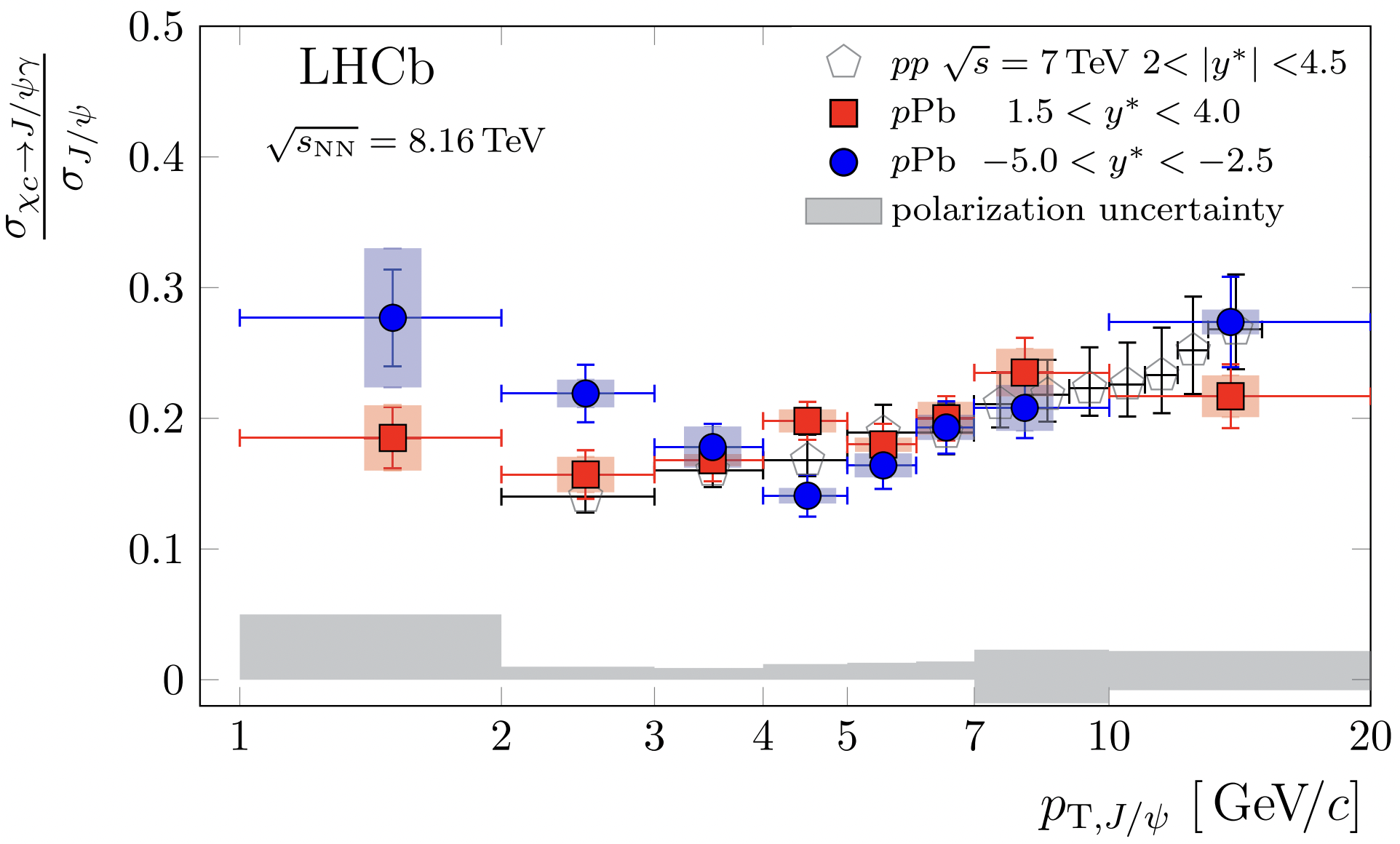}}
\end{minipage}
\hfill\begin{minipage}{0.45\linewidth}
\centerline{\includegraphics[width=1.15\linewidth]{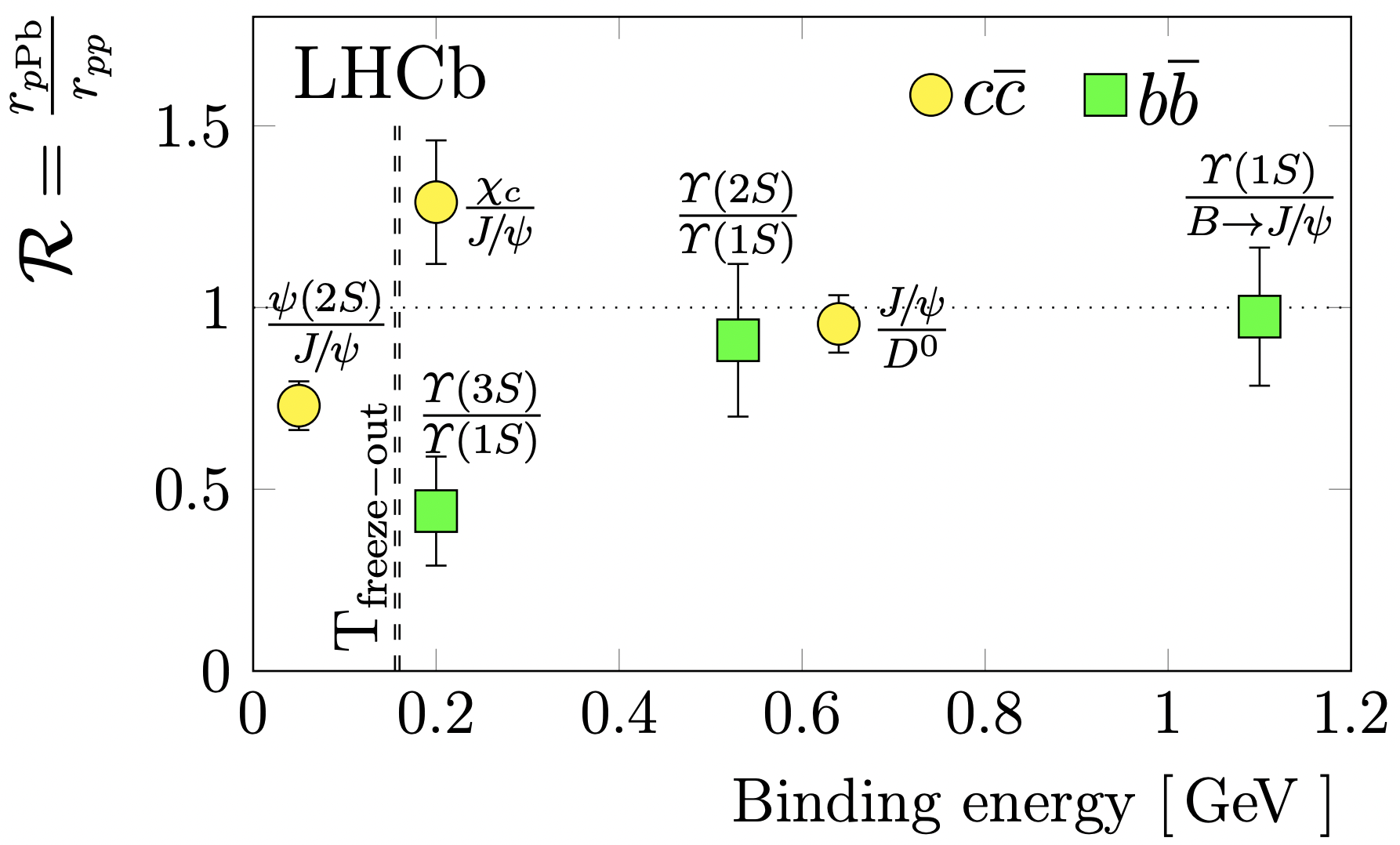}}
\end{minipage}
\caption[]{Left:  Ratio of $\chi_c$ to $J/\psi$ versus $p_{T}$ in 8.16 TeV $p$Pb collisions at forward (red data points) and backward (blue data points) rapidity is compared to the same ratio in 7 TeV $p$$p$ collisions~\cite{results4}.  Right:  Collection of LHCb measurements of excited to ground state quarkonia ratios in $p$Pb collision systems as a function of quarkonia binding energy~\cite{results4}. }
\label{fig:results4}
\end{figure}

The $\sigma_{\Lambda^{0}_{b}}$ to $\sigma_{B^{0}}$ ratio in high multiplicity $p$$p$ collisions~\cite{results2} is shown in Figure~\ref{fig:results2}. The ratio is measured as a function of normalized charged particle VELO tracks (left) and as a function of $p_{T}$ (right). The LHCb measurements are shown by the blue data points, and are compared with $e^{+}e^{-}$ results from LEP~\cite{results22}.  The $\Lambda^{0}_{b}$ to $\sigma_{B^{0}}$ ratio converges at low multiplicity with the LEP result. In the right figure, the enhancement seen at low $p_T$ is inconsistent with PYTHIA calculations, but is well described by EPOSHQ$+$coalescence~\cite{results2a} and the Statistical Hadronization Model~\cite{results2b}, where coalescence refers to $q\bar{q}$ pairs close in phase space that recombine in the later stages of the collision to form mesonic or, in this case, baryonic states.  

\section{Results from $p$Pb collision data}
Figure~\ref{fig:results3} shows the double ratio of the prompt  $\psi(2S)$ to $J/\psi$ cross section as a function of rapidity~\cite{results3}.  The double ratio compares the $\psi(2S)$ to $J/\psi$ ratio in $p$Pb collisions with the same measurement in $p$$p$ collisions.  The prompt contribution shows suppression, where the ratio is slightly below unity and all global errors cancel.  However, the nonprompt double ratio in $p$Pb collisions, although with larger uncertainties, is consistent with unity.  These results suggest a denser nuclear medium could be created in 8 TeV $p$Pb collisions versus 7 TeV $p$$p$ collisions.

A second quarkonia-related measurement is shown in Figure~\ref{fig:results4}, with the ratio of $\chi_c$ to $J/\psi$ as a function of $p_T$ (left) and of quarkonia binding energy (right).  From the left figure, a similar trend is seen above $\sim3$ GeV/c in 8.16 TeV $p$Pb collisions at forward (red data points) and backward (blue data points) rapidity compared to the ratio in 7 TeV $p$$p$ collisions, suggesting no additional nuclear effects are seen in the denser $p$Pb collision system for the more loosely bound $\chi_c$ state.  In the figure on the right, a collection of LHCb measurements of excited to ground state quarkonia ratios are shown in $p$Pb collisions as a function of quarkonia binding energy.  Most of the ratios above the freeze-out temperature (vertical line), are around unity, consistent with the picture of a co-moving nuclear medium as opposed to quark-gluon plasma~\cite{results4}.  

\begin{figure}
\begin{minipage}{0.45\linewidth}
\centerline{\includegraphics[width=1.05\linewidth]{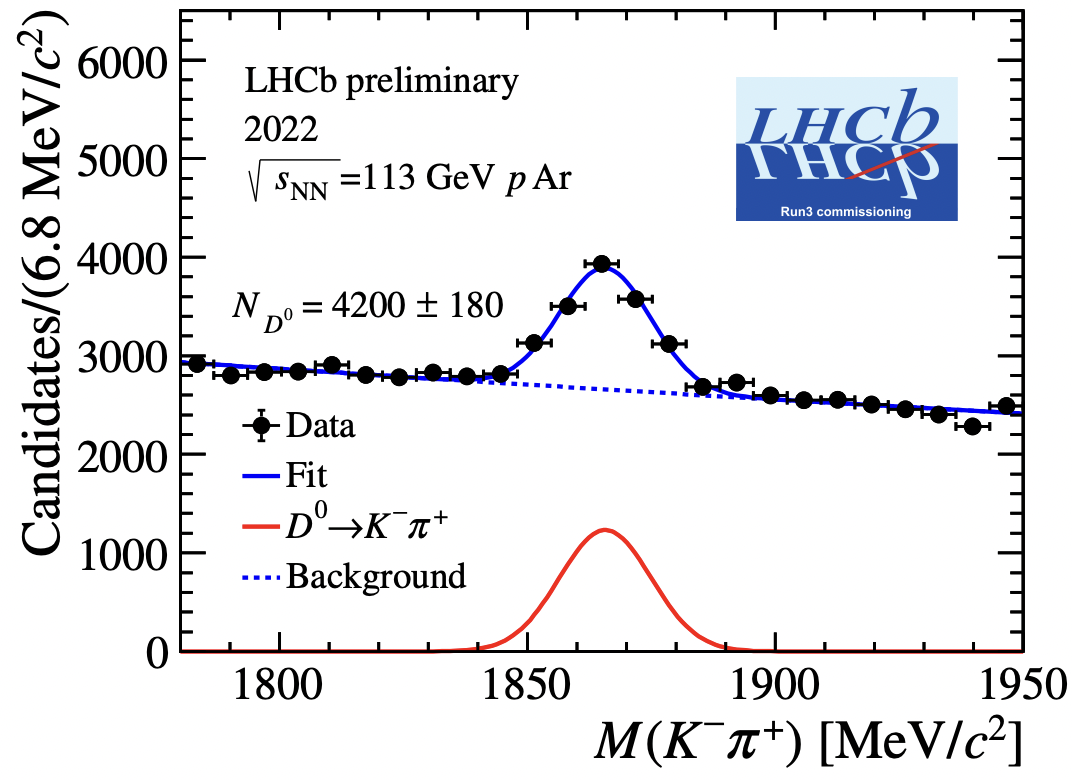}}
\end{minipage}
\hfill\begin{minipage}{0.45\linewidth}
\centerline{\includegraphics[width=1.05\linewidth]{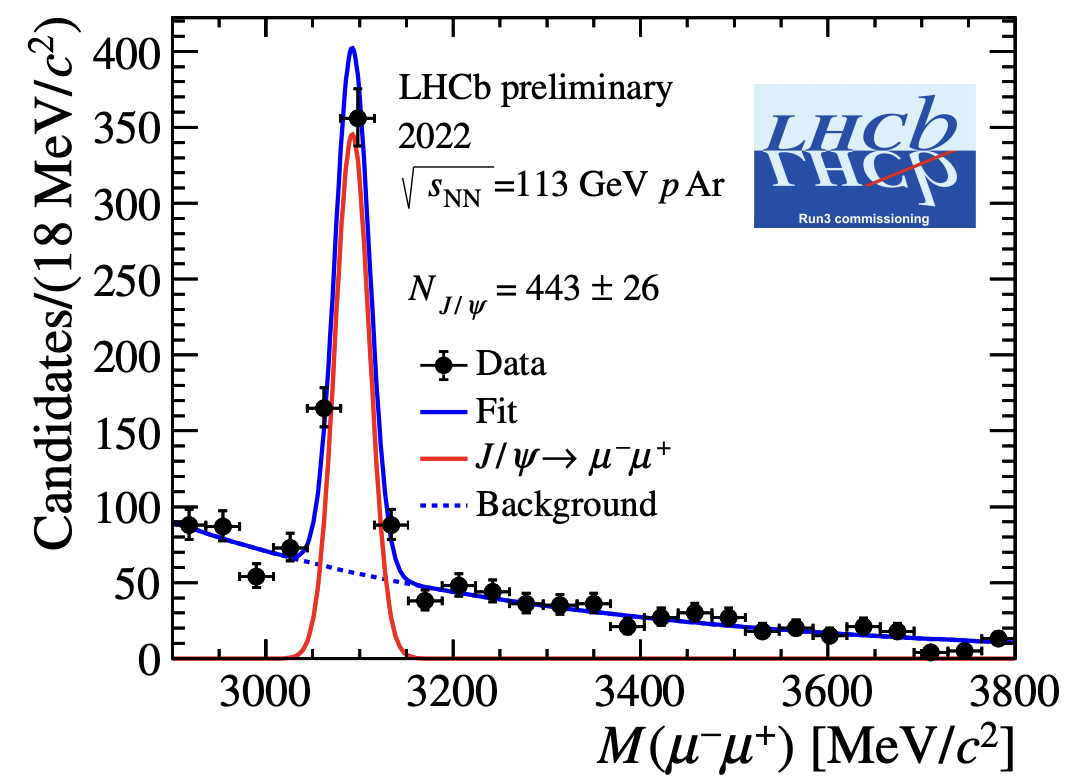}}
\end{minipage}
\caption[]{Run3 initial data from SMOG2~\cite{results6} with only 18 minutes of $p$Ar collisions data taking.  The signal peaks $D^{0} \rightarrow K^{-}\pi^{+}$, left, and $J/\psi \rightarrow \mu^{-}\mu^{+}$, right, are shown~\cite{results6ab}
.}
\label{fig:results6}
\end{figure}

\section{Results from SMOG2 and PbPb collision data}

\begin{wrapfigure}{l}{0.5\linewidth}
\includegraphics[width=\linewidth]{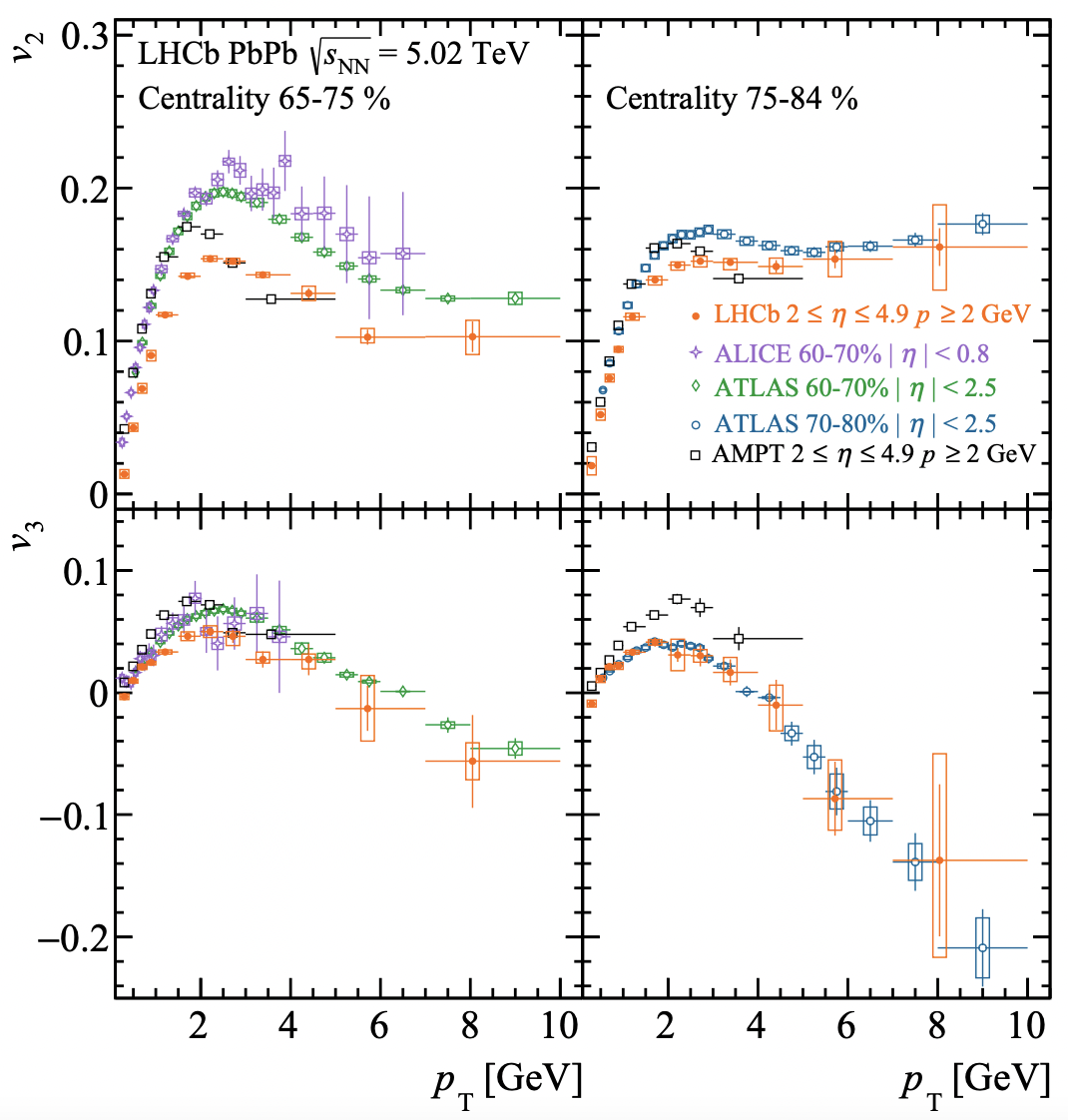}
\caption[]{\label{fig:results5}
First LHC measurements of flow harmonic coefficients $v_2$ and $v_3$ of charged hadrons at forward rapidity in PbPb collisions~\cite{results5} as a function of $p_{T}$.  LHCb results (orange data points) are compared with ATLAS (green and blue data points) and ALICE (violet data points).}
\end{wrapfigure}
The LHCb experiment can simultaneously record data in collider and fixed-target mode by injecting gas in the LHC beam-pipe with SMOG (System for Measuring Overlap with Gas).  Run3 initial data from the upgraded SMOG2 system~\cite{results6} is shown in Figure~\ref{fig:results6} with only eighteen minutes of $p$Ar collision data taking.  The signal peaks $D^{0} \rightarrow K^{-}\pi^{+}$, left, and $J/\psi \rightarrow \mu^{-}\mu^{+}$, right, can clearly be seen in the $\sqrt{s_{_{NN}}}$=113 GeV data~\cite{results6ab}.  Plans for future gas targets include $^{4}$He, $^{20}$Ne, $^{40}$Ar, $^{84}$Kr, $^{132}$Xe, H$_{2}$, N$_{2}$, and O$_{2}$. 

In Figure~\ref{fig:results5}, the first LHC measurements of $v_2$ and $v_3$ of charged hadrons at forward rapidity in PbPb collisions are shown~\cite{results5}.  The LHCb results (orange data points) are compared with ATLAS (green and blue data points) and ALICE (violet data points), where all three measurements follow a similar trend with rising $v_2$ and $v_3$ at low $p_{T}$ that begin to fall approaching higher $p_T$.  This trend is seen in all distributions except $v_2$ for peripheral collisions, potentially due to nonflow effects.  Overall, stronger flow is observed for the ALICE and ATLAS measurements.

\section{Conclusion}
In summary, the LHCb IFT working group has presented new evidence for coalescence with $\Lambda^{0}_{b}$ to $B^{0}$ enhancement in high multiplicity $p$$p$ collisions (selected for Featured in Physics by \textit{Phys. Rev. Lett.}).  The first-ever flow measurements of charged hadrons in PbPb collisions from LHC at forward rapidity have also been presented.  Since the Moriond QCD conference in 2023, there have been new LHCb results in $p$Pb collisions at 8.16 TeV collision energy, including prompt and nonprompt $\psi(2S)$ double ratios and $\chi_c$ production.  Additionally, the $\psi(2S)/J/\psi$ ratio shows suppression with increasing multiplicity in 13 TeV $p$$p$ collisions, consistent with final state effects.  Lastly, SMOG2 results will be coming soon in $p$Ar collisions at 113 GeV center of mass energy, and future plans include fixed-target mode with Nitrogen, Oxygen and more. 



\end{document}